\documentclass{webofc}
\usepackage[varg]{txfonts}   
\usepackage{amsmath}
\usepackage{slashed}
\begin{document}
\title{Short-distance constraints on the hadronic light-by-light}
%
%

\author{\firstname{Johan} \lastname{Bijnens}\inst{1}\fnsep\thanks{\email{johan.bijnens@thep.lu.se}} \and
        \firstname{Nils} \lastname{Hermansson-Truedsson}\fnsep\inst{1,2}\fnsep\thanks{\email{nils.hermansson-truedsson@thep.lu.se}, \emph{Speaker}} \and
        \firstname{Antonio} \lastname{Rodr\'{i}guez-S\'{a}nchez}\inst{3}\fnsep\thanks{\email{arodrigu@sissa.it}}
}

\institute{Department of Astronomy and Theoretical Physics, Lund University, S\"{o}lvegatan 14 A, 223 62 Lund, Sweden
\and
           Albert Einstein Center for Fundamental Physics, Institute for Theoretical Physics, Universit\"{a}t Bern, Sidlerstrasse 5, 3012 Bern, Switzerland
\and
Scuola Internazionale Superiore di Studi Avanzati (SISSA), Via Bonomea 265, 34136, Trieste, Italy
          }

\abstract{%
  The muon anomalous magnetic moment continues to attract interest due to the potential tension between experimental measurement~\cite{Muong-2:2006rrc,Muong-2:2021ojo} and the Standard Model prediction~\cite{Aoyama:2020ynm}. The hadronic light-by-light contribution to the magnetic moment is one of the two diagrammatic topologies currently saturating the theoretical uncertainty. With the aim of improving precision on the hadronic light-by-light in a data-driven approach founded on dispersion theory~\cite{Colangelo:2015ama,Colangelo:2017fiz}, we derive various short-distance constraints of the underlying correlation function of four electromagnetic currents. Here, we present our previous progress in the purely short-distance regime and current efforts in the so-called Melnikov-Vainshtein limit. 
}
\maketitle

\section{Introduction}\label{sec:introduction}
The muon anomalous magnetic moment, or, $a_\mu = (g-2)_\mu /2$, is a potential probe for new physics beyond the Standard Model (SM)~\cite{Aoyama:2020ynm}. Comparing the most recent SM prediction from the White Paper~\cite{Aoyama:2020ynm} with the experimental average from Brookhaven National Laboratory~\cite{Muong-2:2006rrc} and Fermilab National Laboratory~\cite{Muong-2:2021ojo} yields a discrepancy of $4.2\sigma $. The corresponding numbers are
\begin{align}
a_\mu ^{\textrm{SM}} = 116\, 591 \, 810 (43) \times 10^{-11} \, ,
\\
a_\mu ^{\textrm{exp}} = 116\, 592 \, 061 (41) \times 10^{-11} \, .
\end{align}
The SM uncertainty is dominated by hadronic contributions, namely the hadronic vacuum polarisation (HVP) and the hadronic light-by-light (HLbL), which can be calculated either in lattice field theory or using a data-driven approach~\cite{Aoyama:2020ynm}. Although the lattice and data-driven predictions for the HLbL agree within uncertainties, there has since the publication of the White Paper arisen a tension for the HVP~\cite{Borsanyi:2020mff}. The updated lattice prediction for the HVP predicts a significantly reduced discrepancy for the muon magnetic moment~\cite{Borsanyi:2020mff}. There is currently much effort in trying to resolve the tension between the two approaches to calculate the HVP. 

\begin{figure}[t!]
	\centering
	\includegraphics[height=0.17\textheight]{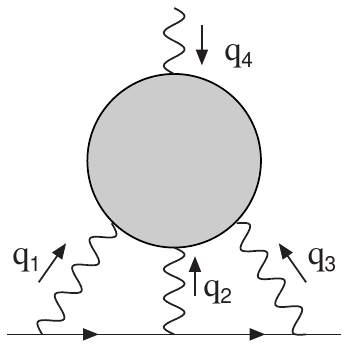}
	\caption{The HLbL topology. The grey blob connecting to the external muon line contains all hadronic contributions. }\label{fig:hlbl}
\end{figure}

In the following, we will be concerned with improving the precision on the HLbL contribution, whose White Paper average is~\cite{Aoyama:2020ynm}
\begin{align}
	a_\mu ^{\textrm{HLbL}} = 92 (18)\times 10^{-11}\, .
\end{align}
Diagrammatically, this contribution is given by the topology in Fig.~\ref{fig:hlbl} where the photons have momenta $q_1$, $q_2$, $q_3$ and $q_4$. The external momentum $q_4$ is for the $(g-2)_{\mu}$ soft, i.e.~in the kinematical limit $q_4\rightarrow 0$.  The remaining three photons have virtual momenta integrated over, which means that in the evaluation of the HLbL there is an intricate mixing between different energy scales. In the dispersive data-driven approach~\cite{Colangelo:2015ama,Colangelo:2017fiz}, one uses analyticity and unitarity to systematically decompose the HLbL into a sum over hadronic contributions. The appearing form-factors in this decomposition require input from e.g.~experiments or the lattice~\cite{Aoyama:2020ynm}.  The unknown hadronic contributions can, however, be controlled with the help of so-called short-distance constraints (SDCs) on the underlying correlation function of four electromagnetic currents, see Refs.~\cite{Melnikov:2003xd,Bijnens:2019ghy,Bijnens:2020xnl,Bijnens:2021jqo}. These kinds of SDCs can be obtained from operator product expansion (OPE) techniques and have been used for model calculations~\cite{Bijnens:1995xf,Colangelo:2019lpu,Colangelo:2019uex,Cappiello:2019hwh,Leutgeb:2019gbz, Colangelo:2021nkr} and a complementary approach building on interpolation techniques~\cite{Ludtke:2020moa}.  

Defining the Euclidean virtualities $Q_{i}^2 = -q_{i}^2$ for $i=1,2,3$, we here consider SDCs in the purely short-distance region $Q_{i}^2\gg \Lambda _{\textrm{QCD}}^2$ (based on Refs.~\cite{Bijnens:2019ghy,Bijnens:2020xnl,Bijnens:2021jqo}), and preliminary work in the Melnikov-Vainshtein limit $Q_{i}^2,Q_{j}^2\gg Q_{k}^2, \Lambda ^2_{\textrm{QCD}}$. These two expansions respectively correspond to three and two of the electromagnetic currents in the underlying correlation function being close. 

\section{Some generalities}

The HLbL is defined in terms of a correlation function of four electromagnetic currents. The corresponding so-called HLbL tensor is given by 
\begin{eqnarray}\label{eq:hlbltensor}
	\Pi^{\mu_{1}\mu_{2}\mu_{3}\mu_{4} } (q_{1},q_{2},q_{3}) = 
	-i\int \frac{d^{4}q_{4}}{(2\pi)^{4}}\left(\prod_{i=1}^{4}\int d^{4}x_{i}\, e^{-i q_{i} x_{i}}\right) 
	\, \langle 0 | T\left(\prod_{j=1}^{4}J^{\mu_{j}}(x_{j})\right)|0\rangle \, .
\end{eqnarray}
Here, the currents are given by $J^{\mu}(x) = \bar{q}\, Q_{q}\gamma ^{\mu}q$ where $q=(u,d,s)$ and $Q_{q}=\textrm{diag}(e_q) =\textrm{diag}(2/3,-1/3,-1/3)$  is the associated charge matrix for light quarks. Here we use the convention $q_{1}+q_{2}+q_{3}+q_{4} = 0$. The above tensor satisfies the Ward identities
\begin{equation}
	q_{i,\, \mu_{i}} \, \Pi^{\mu_{1}\mu_{2}\mu_{3}\mu_{4}}(q_{1},q_{2},q_{3})=0  \, .
\end{equation}
This implies the relation~\cite{Aldins:1970id}
\begin{equation}\label{eq:widerres}
	\hspace{-15pt}
	\Pi^{\mu_{1}\mu_{2}\mu_{3}\mu_{4}}(q_{1},q_{2},q_{3})=-q_{4,\, \nu_{4}}\frac{\partial \Pi^{\mu_{1}\mu_{2}\mu_{3}\nu_{4}}}{\partial q_{4,\, \mu_{4}}}(q_{1},q_{2},q_{3}) \, ,
\end{equation}
which means that one can access $a_{\mu}^{\textrm{HLbL}}$ from the derivative tensor on the right-hand side. The derivative can be Lorentz decomposed into a set of 54 scalar functions $\Pi _{i}$~\cite{Colangelo:2015ama,Colangelo:2017fiz}. The contribution to the magnetic moment can be written in terms of six linear combinations $\hat{\Pi}_{1,4,7,17,39,54}$ of the $\Pi _{i}$, namely
\begin{eqnarray}\label{eq:amuhlblint}
	\hspace{-5pt}
	a_{\mu}^{\mathrm{HLbL}} = \frac{2\alpha ^{3}}{3\pi ^{2}} 
	\int _{0}^{\infty} dQ_{1}\int_{0}^{\infty} dQ_{2} \int _{-1}^{1}d\tau \, \sqrt{1-\tau ^{2}}\,
	 Q_{1}^{3}Q_{2}^{3} \sum _{i=1}^{12} T_{i}(Q_{1},Q_{2},\tau)\, \overline{\Pi}_{i}(Q_{1},Q_{2},\tau)\, ,
\end{eqnarray}
where $\overline{\Pi}_{i}$ are functions of the six $\hat{\Pi}_{j}$, and the kernels $T_i$ are known. Note that the integration in~(\ref{eq:amuhlblint}) is for the full HLbL contribution, whereas the kinematic regions for the two types of SDCs considered here correspond to restricted integration domains. Common for the two limits $Q_{i}^2\gg \Lambda _{\textrm{QCD}}^2$ and $Q_{i}^2,Q_{j}^2\gg Q_{k}^2, \Lambda ^2_{\textrm{QCD}}$ is the need to define an onset $Q_{\textrm{min}}$ of the asymptotic limits: $Q_{i}^2>Q_{\textrm{min}}^2 $ and $Q_{i}^2,Q_{j}^2 > Q_{\mathrm{min}}^2$, respectively. Since there currently is no preferred choice of $Q_{\textrm{min}}$ we keep it variable, and note that this ambiguity induces a source of uncertainty in $a_{\mu}^{\mathrm{HLbL}}$. 

\section{An OPE for three currents}
In the purely short-distance limit $Q_{i}^2\gg \Lambda _{\textrm{QCD}}^2$, three of the currents in the four-point function defined in~(\ref{eq:hlbltensor}) are close. Naively performing an OPE of these currents works only to leading order, since the next-to-leading order contribution in the systematic expansion is ill-defined in the static limit $q_4\rightarrow 0$, which is the limit relevant for the magnetic moment~\cite{Bijnens:2019ghy}. The problem arises from propagators of the momentum $q_4$ which clearly diverge. However, it was shown in Ref.~\cite{Bijnens:2019ghy} that the problem can be circumvented by treating the static photon as an external electromagnetic field and instead of the four-point function $	\Pi^{\mu_{1}\mu_{2}\mu_{3}\mu_{4} }$ consider a three-point correlator in the presence of that field. For this reason we study
\begin{eqnarray}
	\label{eq:3pointem}
	\Pi ^{\mu_{1} \mu_{2} \mu_{3} }(q_{1},q_{2}) = 
	-\frac{1}{e}\int\frac{d^4 q_{3}}{(2\pi)^4}  \left(\prod_{i=1}^{3}\int d^{4}x_{i}\, e^{-i q_{i} x_{i}}\right) \,   \langle 0 | T\left(\prod_{j=1}^{3}J^{\mu_{j}}(x_{j})\right) | \gamma(q_4) \rangle \, ,
\end{eqnarray}
where the static background field is captured in the external state. This is related to the full HLbL tensor through~\cite{Bijnens:2019ghy}
\begin{align}\label{eq:3pt4ptrel}
	\Pi^{\mu_1 \mu_2 \mu_3}(q_{1},q_{2})  = \epsilon_{\mu_4}(q_4)\, \Pi^{\mu_1\mu_2\mu_3\mu_4} (q_{1},q_{2},q_{3}) 
	= -\epsilon_{\mu_4}(q_4)\, q_{4,\, \nu_{4}}\frac{\partial \Pi^{\mu_{1}\mu_{2}\mu_{3}\nu_{4}}}{\partial q_{4,\, \mu_{4}}}
	\, (q_{1},q_{2},q_{3}) \, ,
\end{align}
where $ \epsilon_{\mu_4}(q_4)$ is the polarisation vector of the external photon and the Ward identity~(\ref{eq:widerres}) was used. By performing a systematic background field OPE between the three currents in~(\ref{eq:3pointem}) then yields the HLbL derivative through~(\ref{eq:3pt4ptrel}). The background field induces non-perturbative condensates in the expansion which are different than those appearing in vacuum OPEs~\cite{Shifman:1978bx}. The condensates appear from diagrammatic contributions to the correlator in~(\ref{eq:3pointem}) where some of the fields are left uncontracted in the Wick expansion. Background field OPEs were first used for nucleon magnetic moments in Refs.~\cite{Balitsky:1983xk,Ioffe:1983ju}, but were in fact later also employed for the electroweak contribution to $a_{\mu}$ in Ref.~\cite{Czarnecki:2002nt}.

\subsection{Leading order perturbative and non-perturbative corrections}
\begin{figure}[t!]
	\centering
	\includegraphics[height=0.25\textheight]{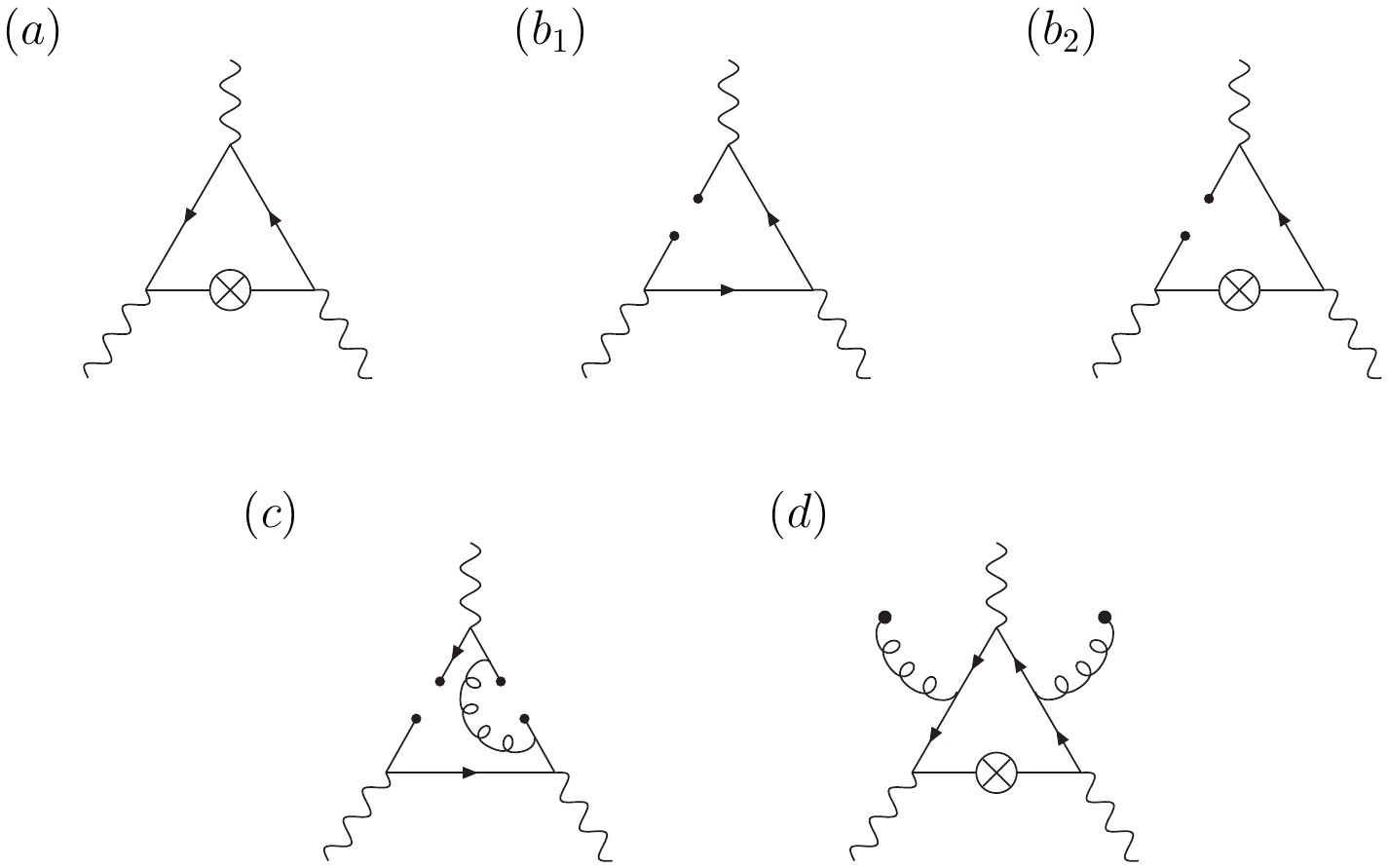}
	\caption{Various contributions to the OPE. The leading order contribution is $(a)$ the quark loop. The crossed vertex indicates the interaction of the external field on the perturbative quark line.  At next-to leading order the non-perturbative condensates $(b_{1,2})$ appear, namely $\langle \overline{q}\, \sigma _{\mu \nu} q \rangle$, induced by $F_{\mu\nu}$, and $\langle \overline{q}q\rangle $. At next-to-next-to leading order diagram $(c)$ is a four-quark operator with condensate $\langle \overline q \, \Gamma _1 q \, \overline{q} \, \Gamma _2 q\rangle$, and diagram $(d)$ contains the gluon condensate $\langle \alpha _s GG\rangle $.}\label{fig:3ptope}
\end{figure}

The systematic OPE is done in powers of $1/Q_i$ and the strong coupling $\alpha _s$. Through order $1/Q_i^6$ and $\alpha _s$ the leading order term and non-perturbative corrections arise from the operators~\cite{Bijnens:2019ghy,Bijnens:2020xnl}
\begin{align}
	S_{1,\,\mu\nu}
&
	=  e\,   e_{q}  F_{\mu\nu} 
	\, ,
\qquad
	S_{2,\,\mu\nu}
	=  \bar{q}\, \sigma_{\mu\nu}q   \, ,
	\hspace{17pt}
 \quad
	S_{3,\,\mu\nu}
	=  i \,  \,\bar{q} \, G_{\mu\nu}q  \, ,
	\\ 
	S_{4,\,\mu\nu}
&	=  i \, \bar{q} \, \bar{G}_{\mu\nu}\gamma_{5} q \, ,
\quad
	S_{5,\,\mu\nu}
	= \bar{q} q\; e\,   e_{q}F_{\mu\nu} \, ,
	\hspace{-10pt}
\qquad
	S_{6,\,\mu\nu}
	=  \frac{\alpha_{s}}{\pi}\, G_ {a}^{\alpha\beta}G^{a}_{\alpha\beta}\; e\,   e_{q}F_{\mu\nu}  \, ,
	\\
	S_{7,\,\mu\nu}
&
	=  \bar{q}\, \left(G_{\mu\lambda}D_{\nu}+D_{\nu}G_{\mu\lambda}\right) \gamma^{\lambda}q-(\mu\leftrightarrow\nu) \, ,
	\\
	S_{\{8\},\,\mu\nu}
&
	=  \alpha_{s}\, \left( \bar{q}\, \Gamma q \; \bar{q}\, \Gamma q\right) _{\mu\nu} \, .
\end{align}
Here, the gluon field is given by $G_{\mu\nu}= i g_S\lambda^a G^a_{\mu\nu}$ as well as its dual $\bar{G}^{\mu\nu} = \frac{i}{2}\epsilon^{\mu\nu\lambda\rho} G_{\lambda\rho}$. The matrix $\Gamma$ is in flavour, spin and colour space . For more detail, we refer the reader to Ref.~\cite{Bijnens:2020xnl}. The class of four-quark operators  $ S_{\{8\},\,\mu\nu}$ yields flavour mixing, and in the chiral limit there are only 12 independent operators. In Fig.~\ref{fig:3ptope} we show the various diagrammatic contributions appearing in the OPE. The leading order term is given by the perturbative quark-loop in Fig.~\ref{fig:3ptope}(a), which in fact coincides with the leading order term in a naive OPE of the four-point function in~(\ref{eq:hlbltensor}). This finding in Ref.~\cite{Bijnens:2019ghy} confirmed that indeed the perturbative quark loop describes the leading behaviour in the short-distance limit~\cite{Aoyama:2020ynm}.

The obtained systematic expansion allows us to quantify for the first time from where the short-distance representation of the HLbL is valid. We study first the impact of the non-perturbative corrections. We note in passing that one also has to renormalise the condensates, which is described in detail in Ref.~\cite{Bijnens:2020xnl}. It suffices here to write the renormalised result on the form
\begin{equation}\label{eq:operesren}
	\hspace{-20pt}
	\Pi^{\mu_{1}\mu_{2}\mu_{3}}(q_{1},q_{2}) = \frac{1}{e}\, \vec{C}^{\, T,\, \mu_{1}\mu_{2}\mu_{3}\mu_{4}\nu_{4}}_{\overline{MS}}(q_{1},q_{2})\, \langle \vec{Q}_{\overline{MS}, \, \mu_{4}\nu_{4}}(\mu)\rangle
	\, ,
\end{equation}
where the perturbative short-distance Wilson coefficients are contained in the vector $ \vec{C}^{\, T,\, \mu_{1}\mu_{2}\mu_{3}\mu_{4}\nu_{4}}_{\overline{MS}}$ and the condensates at renormalisation scale $\mu$ in $\langle \vec{Q}_{\overline{MS}, \mu_{4}\nu_{4}}(\mu)\rangle$. For convenience, we write 
\begin{equation}
	\label{eq:magnsusc}
	\langle\vec{Q}_{\overline{MS}, \, \mu\nu}(\mu)\rangle =
	e\, \vec{X} \, \langle e_{q}F_{\mu\nu}\rangle \, .
\end{equation} 
Numerical estimates for the condensates are obtained from Refs.~\cite{Bijnens:2020xnl,Shifman:1978bx,Aoki:2019cca,Belyaev:1982sa}. 

\begin{figure*}[t!]
	\begin{minipage}{.49\textwidth}
		\centering
		\includegraphics[width=1.\textwidth]{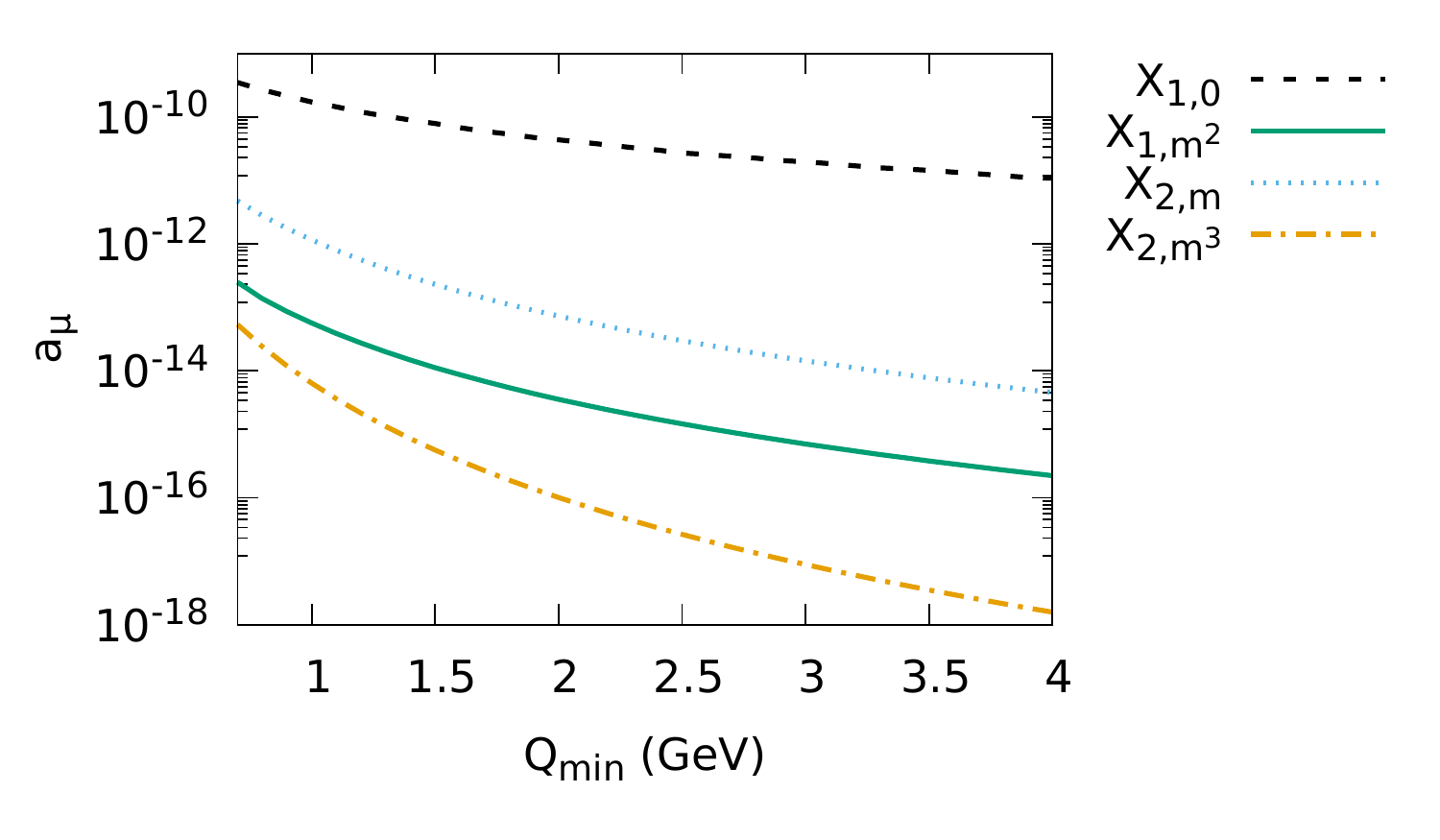}
		\caption{Contributions from $X_{1,0}$, $X_{1,m^2}$, $X_{2,m}$ and $X_{2,m^3}$ to $a_{\mu}^{\mathrm{HLbL}}$. }
		\label{fig:plot1}
	\end{minipage}
\hspace{0.5cm}
	\begin{minipage}{.49\textwidth}
		\centering
		\includegraphics[width=1.\textwidth]{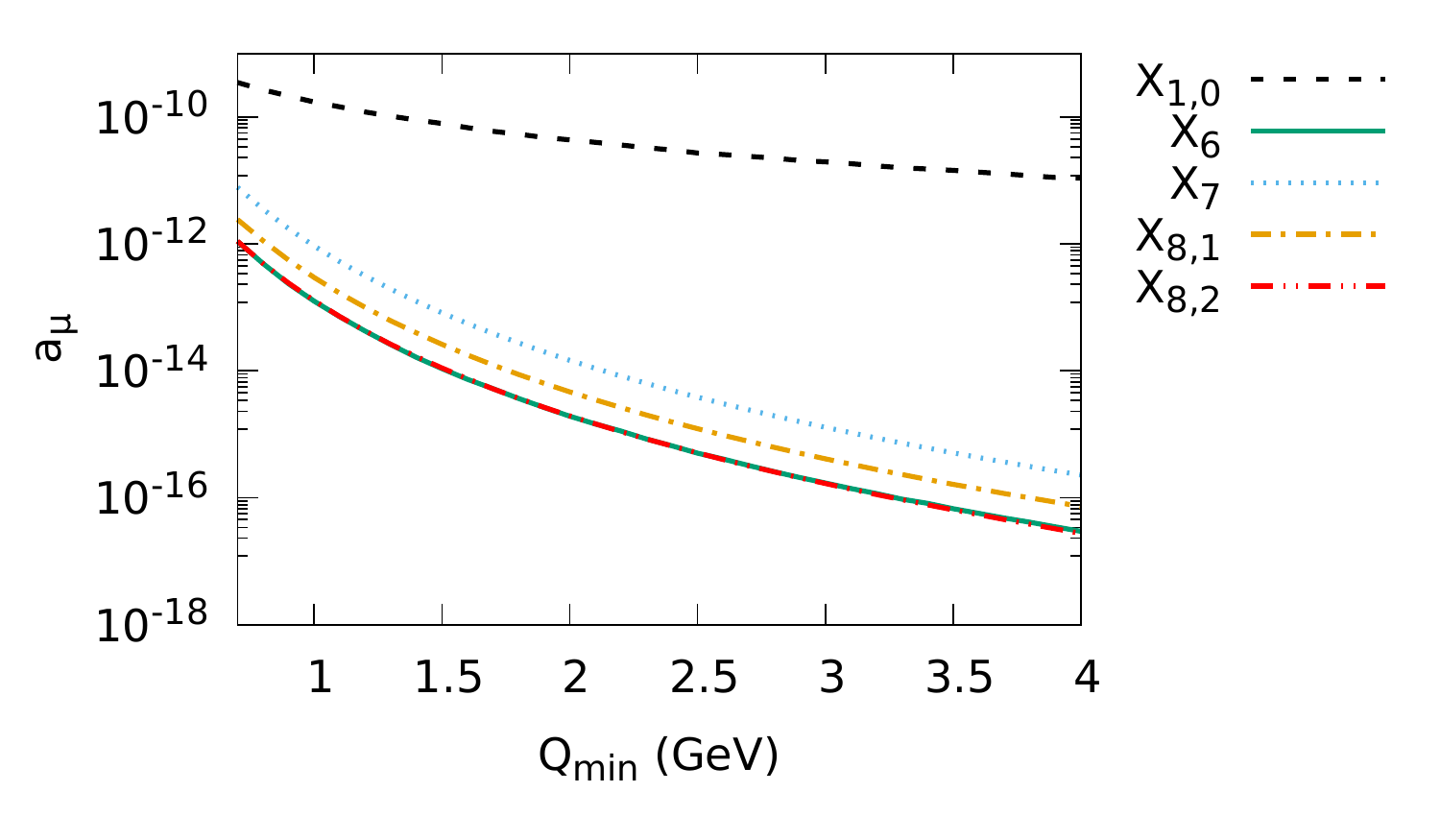}
		\caption{Contributions from $X_{1,0}$, $X_{6}$, $X_{7}$, $X_{8,1}$ and $X_{8,2}$ to $a_{\mu}^{\mathrm{HLbL}}$. }  
		\label{fig:plot2} 
	\end{minipage}
\end{figure*}

\begin{figure*}[t!]
	\centering
	\centering \includegraphics[width=0.49\textwidth]{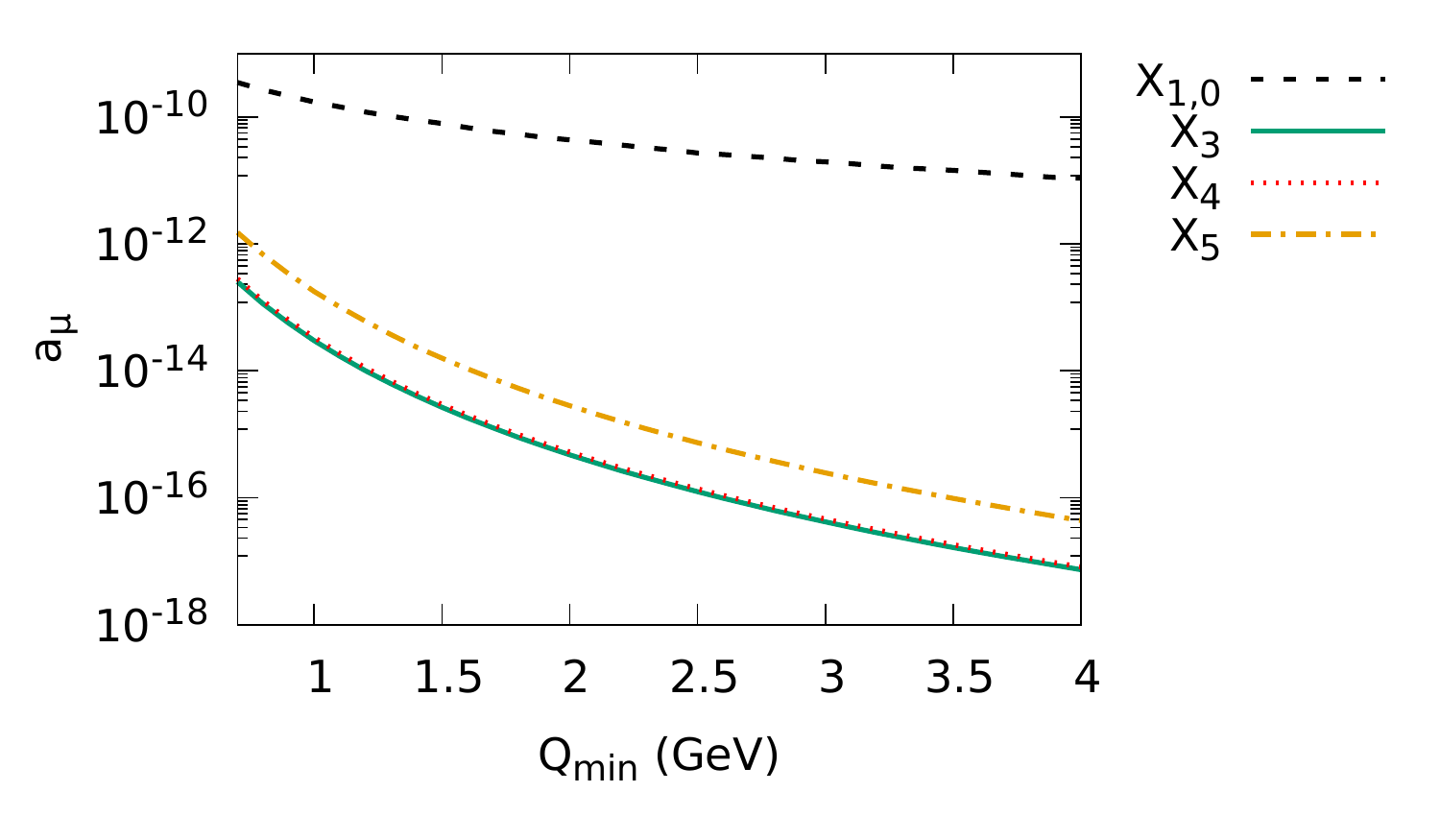}
	\caption{Contributions from $X_{1,0}$, $X_{3}$, $X_{4}$ and $X_{5}$ to $a_{\mu}^{\mathrm{HLbL}}$. }
	\label{fig:plot3}
\end{figure*}

Analytical formulae for the $\hat{\Pi}_{i}$ needed  to numerically calculate $	a_{\mu}^{\mathrm{HLbL}} $ in~(\ref{eq:amuhlblint}) in the short-distance region $Q_{i}^2 > Q_{\textrm{min}}^2$ can be found in Refs.~\cite{Bijnens:2019ghy,Bijnens:2020xnl}. Performing the numerical integration yields Figs.~\ref{fig:plot1}--\ref{fig:plot3}.  The leading order perturbative quark loop corresponds to $X_{1,0}$, which as can be seen is completely dominant by one to two orders of magnitude. This shows that the non-perturbative corrections through next-to-next-to leading order are negligible as compared to the perturbative quark loop. 


\subsection{Gluonic corrections}
\begin{figure*}[t!]
	\begin{minipage}{.49\textwidth}
		\centering
	\includegraphics[height=0.12\textheight]{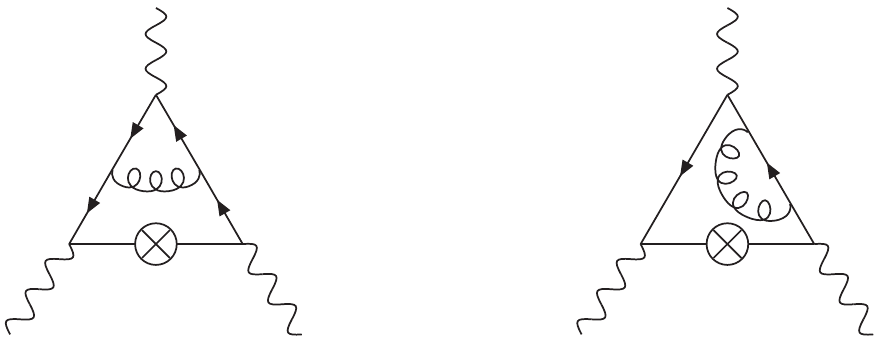}
	\caption{Two-loop gluonic corrections to the perturbative quark loop.  }\label{fig:gluondiagrams}
	\end{minipage}
	\hspace{0.5cm}
	\begin{minipage}{.49\textwidth}
	\centering
\includegraphics[height=0.2\textheight]{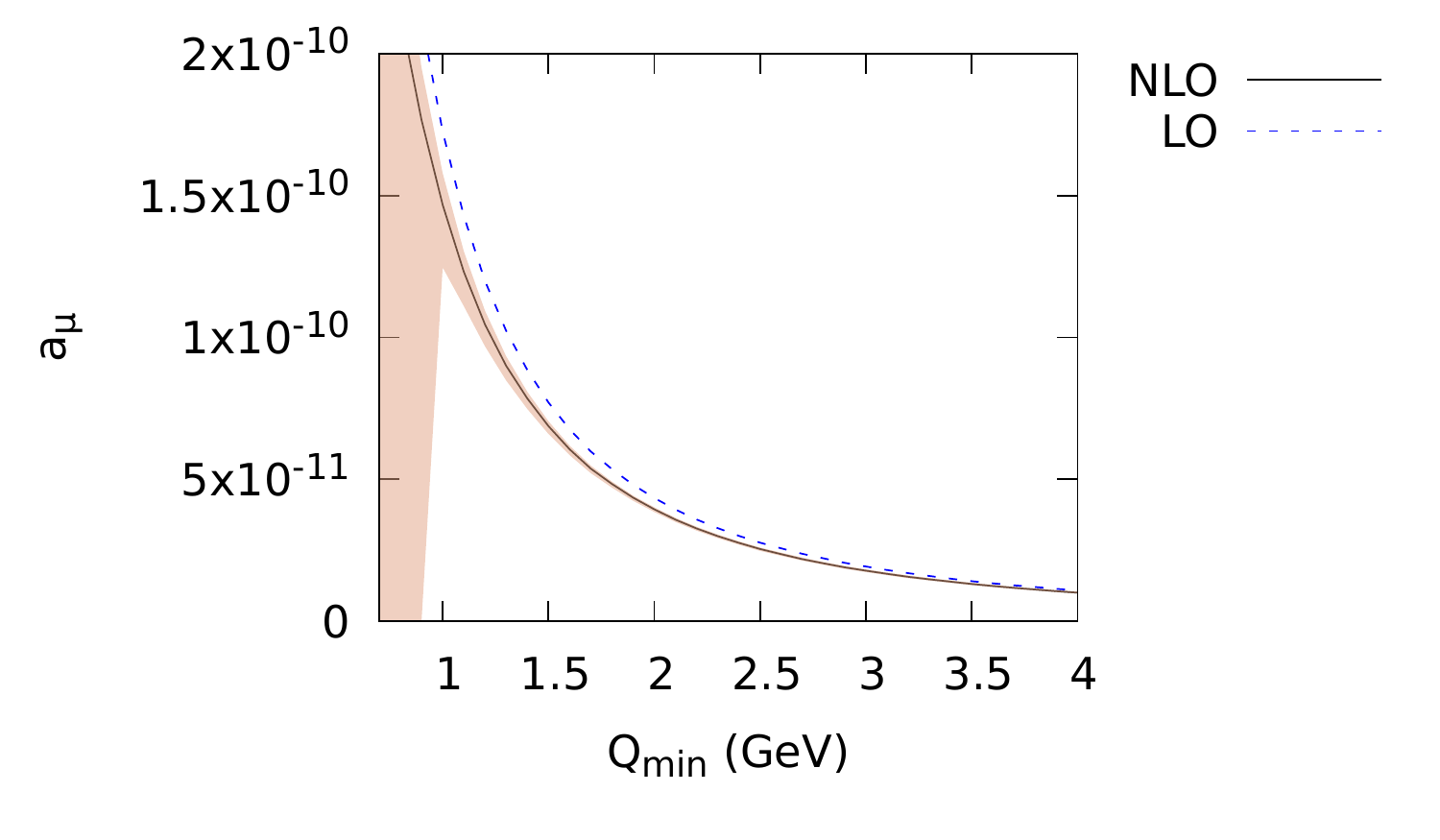}
\caption{Contributions to $a_{\mu}^{\mathrm{HLbL}}$ from  gluonic corrections to the perturbative quark loop. The uncertainty band is associated to $\alpha _{s}\left( \mu = Q_{\textrm{min}}\right) $.  }\label{fig:gluonnum}
	\end{minipage}
\end{figure*}

From the above numerical results it appears that the perturbative quark loop is a good description of the short-distance behaviour of the HLbL. However, perturbative order $\alpha _{s}$ corrections to the quark loop can also be important, which is what we studied in Ref.~\cite{Bijnens:2021jqo}. Including two vertices from the Dyson series with gluon interactions yields two-loop topologies like those in Fig.~\ref{fig:gluondiagrams}. Details of the calculation are given in Ref.~\cite{Bijnens:2021jqo} together with the $\hat{\Pi}_i$, and we report here only the numerical comparison to the quark loop contribution to $	a_{\mu}^{\mathrm{HLbL}} $ in Fig.~\ref{fig:gluonnum}. As can be seen, the shift in going from the leading order quark loop to including the two-loop corrections is small. Numerically, it corresponds to a shift around $-10\%$ compared to the quark loop. This indeed shows that the quark loop describes the short-distance dynamics of the HLbL to within 10\% in the region $Q_{i}^2 \gg \Lambda  _{\textrm{QCD}}^2$, and can be used by dispersive and model studies as in Ref.~\cite{Colangelo:2021nkr}.

\begin{figure}[t!]
\end{figure}

\section{An OPE for two currents}
Having concluded the study for the purely short-distance region $Q_{i}^2 \gg \Lambda  _{\textrm{QCD}}^2$ we now turn to the Melnikov-Vainshtein limit $Q_{i}^2,Q_{j}^2\gg Q_{k}^2, \Lambda ^2_{\textrm{QCD}}$. This was studied first in Ref.~\cite{Melnikov:2003xd}. We emphasise that results here are still preliminary, and we exemplify by considering the case $Q_{1}^2,Q_{2}^2\gg Q_{3}^2, \Lambda ^2_{\textrm{QCD}}$. Recall that for three currents we studied the three-point function in~(\ref{eq:3pointem}). In the Melnikov-Vainshtein limit we can instead use the two-point function
\begin{equation}\label{eq:2ptem}
	\Pi^{\mu_{1}\mu_{2}}=\frac{i}{e^{2}}\int \frac{d^{4}q_4}{(2\pi)^4}\int d^{4}x_1\int d^{4}x_2 e^{-i(q_1 x_1+q_2 x_2)}\langle 0 |T \left\{ J^{\mu_1}(x_1)J^{\mu_2}(x_2)\right\} |\gamma ^{*}(q_3) \gamma(q_4) \rangle \, .
\end{equation}
The relation to the four-point HLbL tensor is then (cf.~(\ref{eq:3pt4ptrel}))
\begin{equation}\label{eq:2pt4ptrel}
	\Pi^{\mu_1\mu_2}=\epsilon_{\mu_3}(q_3)\, \epsilon_{\mu_4}(q_4) \Pi^{\mu_1\mu_2\mu_3\mu_4}  = -\epsilon_{\mu_3}(q_3)\,  \epsilon_{\mu_4}(q_4)\, q_{4,\, \nu_{4}}\frac{\partial \Pi^{\mu_{1}\mu_{2}\mu_{3}\nu_{4}}}{\partial q_{4,\, \mu_{4}}}\, (q_{1},q_{2},q_{3}) \, .
\end{equation}
In this OPE we will expand in powers of the large momentum $\hat{q} = (q_1-q_2)/2$. We further define a Dirac matrix in terms of the massless quark propagator according to
\begin{align}
	\Gamma ^{\mu \nu}(k) = \gamma ^{\mu}S(k)\gamma^{\nu} \, , \qquad S(k) = \frac{\slashed{k}}{k ^2} \, .
\end{align}
Denoting the perturbative quark loop contribution to the HLbL tensor in~(\ref{eq:hlbltensor}) by $\Pi^{\mu_1\mu_2\mu_3\mu_4}_{\mathrm{quark-loop}}$, one finds in the chiral limit through next-to-leading order the result
\begin{align}\label{eq:2ptope}
	\Pi^{\mu_{1}\mu_{2}}
	&\approx {- \frac{ e_q^2}{e^2}\,   \left\langle \bar{q}(0)\left[\Gamma^{\mu_1\mu_2}(-\hat{q})-\Gamma^{\mu_2\mu_1}(-\hat{q})\right] q(0)\,  \big| \gamma^{*}(q_3) \gamma (q_4) \right\rangle 
	}
\nonumber 
\\
&
-\frac{ie_q^2}{e^2\hat{q}^2}\left(
g^{\mu_1\delta}g^{\mu_2}_ {\beta}
+g^{\mu_2\delta}g^{\mu_1}_{\beta}
-g^{\mu_1\mu_2}g^{\delta}_{\beta}\right)
\left(g_{\alpha\delta}-2\frac{\hat{q}_\delta \hat{q}_\alpha}{\hat{q}^2}\right)
  \left\langle \bar{q}(0)\left[ \overrightarrow{\partial}^{\alpha}-\overleftarrow{\partial}^{\alpha}\right] \gamma^{\beta}  q(0) \big| \gamma^{*}(q_3) \gamma (q_4)\,  \right\rangle 
\nonumber
\\
&
-\frac{1}{4}{F_{\nu_3\mu_3}F_{\nu_4 \mu_4}}\left. \frac{\partial}{\partial q_{3\, \nu_3}}\right| _{q_3 \rightarrow 0}
\left. 
\frac{\partial}{\partial q_{4 \, \nu_4}}\right|_{q_4 \rightarrow 0}\Pi^{\mu_1\mu_2\mu_3\mu_4}_{\mathrm{quark-loop}} \, .
\end{align}
We thus see that the two external photons in~(\ref{eq:2ptem}) induce new contributions as compared to~(\ref{eq:3pointem}). There are several technical issues in obtaining the $\hat{\Pi}_i$ in the Melnikov-Vainshtein limit which will be explained in our upcoming paper. Note that we have made no ordering on the size of $Q_{3}^2$ and $\Lambda _{\mathrm{QCD}}^2$ here, except for an omitted gluon operator contribution to the flavour singlet piece when $\Lambda_{\mathrm{QCD}}<Q_3$. However, the perturbative $Q_{3}^2$ regime $Q_{1}^2,Q_{2}^2\gg Q_{3}^2 \gg \Lambda ^2_{\textrm{QCD}}$ corresponds to a special case of the purely short-distance limit $Q_{i}^2 \gg \Lambda  _{\textrm{QCD}}^2$ discussed above. This means that for $Q_{3}^2 \gg \Lambda ^2_{\textrm{QCD}}$ in the Melnikov-Vainshtein limit we should reproduce our previous results for the perturbative quark loop as well as the leading order result of Ref.~\cite{Melnikov:2003xd}. Evaluating the appearing matrix elements in~(\ref{eq:2ptope}) perturbatively, this is indeed the case. 

In the non-perturbative region $Q_{3}^2 \slashed{\gg}  \Lambda ^2_{\textrm{QCD}}$, the matrix elements in~(\ref{eq:2ptope}) must be form-factor decomposed. The leading order term (first row in~(\ref{eq:2ptope})) depends on two form-factors, i.e.~the longitudinal and transversal functions discussed in Ref.~\cite{Melnikov:2003xd}. At next-to-leading order there are six form-factors, and the exact decomposition will be given in our future paper. 

We are currently working on gluonic corrections to the OPE in~(\ref{eq:2ptope}). As argued in Ref.~\cite{Ludtke:2020moa}, the leading perturbative $\alpha _s$ correction is expected to be given by the Melnikov-Vainshtein result~\cite{Melnikov:2003xd} with an overall factor $-\alpha _{s}/\pi$. In our OPE, through leading order in $\alpha_s$ the first line in~(\ref{eq:2ptope}) instead yields
\begin{align}\label{eq:2ptopegluon}
	\Pi^{\mu_{1}\mu_{2}}_{\mathrm{LO},\, \alpha _s}
	&\approx {- \frac{ e_q^2}{e^2}\, \left( 1-\frac{7}{3}\, \frac{\alpha _s}{\pi}\right) \,   \left\langle \bar{q}(0)\left[\Gamma^{\mu_1\mu_2}(-\hat{q})-\Gamma^{\mu_2\mu_1}(-\hat{q})\right] q(0)\,  \big| \gamma^{*}(q_3) \gamma (q_4) \right\rangle 
	}
 \, .
\end{align}
Although this might seem to be in tension, the Melnikov-Vainshtein result refers to the axial current that preserves chiral symmetry within perturbation theory, related to ours by a finite counterterm~\cite{Trueman:1979en}.  We are currently working to complete the calculation, taking these subtleties into account. 


SDCs in the Melnikov-Vainshtein limit are highly relevant for data-driven approaches to evaluate the HLbL, see e.g.~Refs.~\cite{Melnikov:2003xd,Colangelo:2019lpu,Colangelo:2019uex,Ludtke:2020moa,Colangelo:2021nkr}. Our anticipated results can be used to further improve the data-driven prediction. 

\section{Conclusions}
In these proceedings we have reported on recent progress on the derivation of short-distance constraints for the data-driven evaluation of the HLbL. This hadronic contribution to the muon magnetic moment is particularly complicated since it mixes long- and short-distance dynamics. Constraints in the purely short-distance as well as the Melnikov-Vainshtein limit are important to control the systematic uncertainty in the data-driven approach to the HLbL, both of which we have considered herein. Due to the soft external magnetic field, the constraints are derived using background field operator product expansion techniques. Our main result is a precise description for the HLbL in the purely short-distance regime. In the Melnikov-Vainshtein limit, which is still work in progress, we have reproduced important results in certain limits, and are currently evaluating also novel higher-order non-perturbative as well as gluonic corrections. 

\section*{Acknowledgements}
N.~H.-T. is funded in part by the Albert
Einstein Center for Fundamental Physics at the University of Bern, and in part by the Swedish Research Council, project number 2021-06638. J.~B.~is supported by the Swedish
Research Council grants contract numbers 2016-05996 and 2019-03779. A.~R.-S.~is funded in part by MIUR contract number 2017L5W2PT.

\bibliography{references}

\end{document}